# Musical consonance: a review of theory and evidence on perception and preference of auditory roughness in humans and other animals

John M. McBride[1]*
[1]Acoustics Research Institute, Austrian Academy of Science, Vienna, Austria

This review paper is still under active development. Peer review and comments are welcome, and should be addressed to John McBride: *jmmcbride@protonmail.com.

## Abstract

The origins of consonance in human music has long been contested, and today there are three primary hypotheses: aversion to roughness, preference for harmonicity, and learned preferences from cultural exposure. While the evidence is currently insufficient to disentangle the contributions of these hypotheses, I propose several reasons why roughness is an especially promising area for future study. The aim of this review is to summarize and critically evaluate roughness theory and models, experimental data, to highlight areas that deserve further research. I identify 2 key areas: There are fundamental issues with the definition and interpretation of results due to tautology in the definition of roughness, and the lack of independence in empirical measurements. Despite extensive model development, there are many duplications and models have issues with data quality and overfitting. Future theory development should aim for model simplicity, and extra assumptions, features and parameters should be evaluated systematically. Model evaluation should aim to maximise the breadth of stimuli that are predicted.

## 1. Introduction

What makes music sound pleasant? This problem has excited researchers for millennia.[1] Early philosophers hypothesized about the "naturalness" of certain sounds due to the simple mathematics that described harmonic intervals.[2,3] Developments from the Renaissance period onward helped us understand the structure and physics of sound.[4] The harmonic theory of consonance was supported by the observation that harmonic intervals could be found in the harmonic series, although we still lacked a genuine scientific explanatory theory for why these intervals should be preferred other than "naturalness". The 19th century saw the birth of modern psychoacoustics and empirical testing, alongside the first genuine scientific theory of consonance.[5] Helmholtz theorised that consonance is the absence of "sensory dissonance", or "roughness", a sensation that is perceived when tones interact to produce "beats" – audible amplitude modulation patterns – at rates of about 15 - 100 Hz. Despite voluminous research on the subject, we still lack a definitive answer on the origins of consonance. This stalemate is perhaps best illustrated by how common it is for papers in this field to be followed by replies and rebuttals.[6] The leading hypotheses are that consonance is due to: (i) learned preferences due to cultural exposure; (ii) aversion to roughness; (iii) preference for harmonicity – harmonic sounds are defined as combinations of complex tones with overlapping harmonics. Other hypotheses of note are that dissonance is caused by "sharpness"[7] – tones with energy in high frequencies – while a neurodynamic theory relates consonance to mode-locking stability of neural synchronization.[8]

The main barriers to understanding the origins of consonance is that these hypotheses can be instantiated in different ways, most of which (except for sharpness) have the same predictions.[9–12] There is a chicken and egg problem, in that culture and (potentially partially innate) preferences can interact over time in a feedback loop. We know that preferences can form purely from repeated exposure, and that the prevalence of intervals in music mirrors human preferences. The prevalence of harmonic intervals in music around the world[13,14] may be due to an innate bias towards preferring certain sounds, or it may be that other factors caused the prevalence and the prevalence then led to preferences.[11–13] The empirical interval preferences and prevalence in Western countries is predicted quite well by models based on the roughness, harmonicity and neurodynamics hypotheses. So we are left with a situation where it is possible that none of these theories lead to innate preferences, but influence prevalence in another way, and prevalence then leads to preferences. Or one or more of these theories does lead to innate preferences, but we cannot distinguish which hypothesis is better. Cross-cultural studies could potentially dissociate interval prevalence from other predictions, but these studies are typically confounded by two factors: globalization and homogenization of musical tonalities, and pre-existing similarities between musical tonalities used in different cultures. Notable exceptions exist where researchers sought people in remote locations that have limited exposure to globalized music, although this approach is costly and difficult to scale.[15–19] Another two recent advances are more tractable, where confounds with cultural prevalence is avoided through the use of intervals and timbres that are not often heard in Western music. Several studies have used the Bohlen-Pierce scale for which many intervals are far from 12-tone equal temperament tuning,[20–23] and others have used dense ratings that sample intervals uniformly over approximately an octave range.[24–26] Other studies have used timbres that are not common in Western music, in particular inharmonic timbres that lead to very different predictions (from harmonicity and roughness models) than standard harmonic tones.[24,25,27] Another important advance is methodological, as new sampling methods have been developed to deal with the problem of large stimulus spaces.[24,26,28] Given these advances, I am quite optimistic about future progress on this age-old question.

In this review I focus on roughness. Although there is yet insufficient evidence to say which theory best explains consonance, I focus on roughness because I think it has the most well-developed theoretical foundations and development, and currently has the strongest cross-cultural empirical support. Ultimately, the goal of this review is to highlight some underlying issues at the core of the roughness hypothesis, and to direct future work. Given the recent advances in sampling of subjects (online experiments, access to remote locations), sampling of subject responses (e.g. Gibbs sampling with people),[28] leading to increased sampling of stimuli space, it feels like the time is right to reevaluate the theoretical foundations and basic empirical findings. Despite decades of work and dozens of papers devoted to developing models, most models are copies of each other, and they all rely on small sets of training data, data that is old and potentially unreliable, and in some cases simulated data instead of empirical data. There is a tendency towards overfitting, and a reliance on weakly tested assumptions. And at the heart of the issue, there is a question about what ‘roughness’ really is, and whether it is a cause of dissonance, or a description of a particular sort of dissonance. The future is bright, but first we should take a step back to re-evaluate what is the best next step forward. Aspects of roughness theory should be re-evaluated given the results of this review, and a renewed attempt should be made at developing and testing models and their assumptions. Hopefully this will spur parallel theoretical advances for the alternate hypotheses. Ultimately, the origins of consonance will be unravelled through theory-driven testing. This review points to the next steps on that path.

In Section 2 I briefly explain the rationale behind my choice of prioritising roughness, alongside an overview of other theories. In Section 3 I describe roughness, and its basis in physics and physiology. In Section 4 I present an overview of roughness models, grouped into two families. In Section 5 I discuss strengths and weaknesses of the modelling approaches, and suggest future directions. In Section 6 I discuss the empirical evidence for perception and preference of roughness (or more simply, amplitude-modulated tones) in humans and non-human animals (hereafter referred to as animals).

## 2. Hypotheses on the Origins of Consonance

To be able to clearly evaluate hypotheses on the origins of consonance it helps to consider them systematically in a logical framework. For this I look to Ernst Mayr's consideration of proximate vs ultimate causes that became popular in evolutionary biology.[29] Proximate causes are considered mechanistic and immediate, while the ultimate causes are considered long-term and historical. However, there is some ambiguity in these definitions since there is no one proximate or ultimate cause, and there may even be loops in the causal chain due to interactions between causes.[30] To illustrate this, consider the chain of hypothetical causes that could lead to consonance preferences in **Fig. 1**.

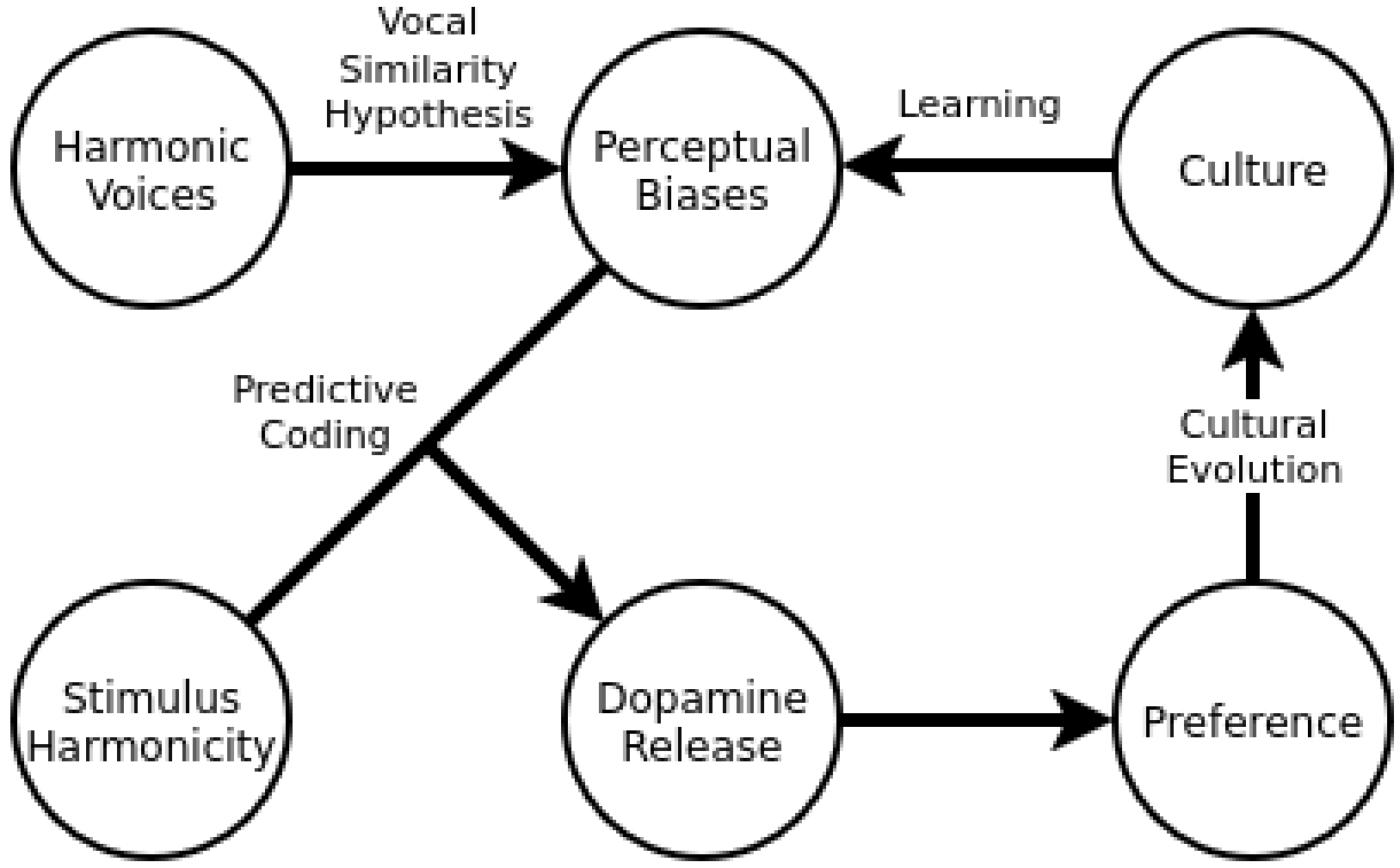

**Figure 1.** At the end of this chain we observe a participant reporting their preference for a sound. According to predictive coding theory the preference rating decision was caused by the amount of dopamine released in the brain. The amount of dopamine released in the brain depends on the 'harmonicity' of the sound, and perceptual biases in the processing of harmonic sounds. These perceptual biases are caused by innate properties of harmonic sounds / auditory processing, and also caused by cultural exposure to harmonic sounds. Culture then leads to increased use of harmonic sounds in a feedback loop. Over even longer timescales, the regularity of harmonicity in human voices leads to the perceptual biases for processing harmonic sounds through evolution.

The above example is a scenario based on the 'vocal similarity hypothesis', and is just one possible scenario out of many. Note how definitions of proximate vs ultimate are ambiguous due to a multiplicity of causes and recursion that allows feedback loops. Due to these issues of ambiguity and recursion, the proximate / ultimate distinction can have limited value. For the following, however, we use the terms to delineate the levels of causation at clearly different timescales. Proximate causes here refer to mechanistic causes that occur over short time spans, as one listens to the music, or as one forms expectations and preferences over a lifetime. Ultimate causes here refer to causes that are not specific to an individual's life trajectory, and instead have evolutionary roots.

## 2.1. Proximate Causes

I give only a brief overview of five hypotheses on proximate causes of consonance: learning, roughness, harmonicity, sharpness, and neurodynamics. Each of these has some evidence for it, and each requires more development. I will highlight some key strengths and weaknesses of each hypothesis.

### *Learning / Culture: We like what is familiar.*

Lived experience tells us that we learn to like. The degree of variation in musical styles (across and within societies) and personal tastes makes it clear that innate perceptual biases do not strictly determine preferences. Some people enjoy listening to music that is disliked by the majority of people (atonal, noise, etc.). Empirical studies have shown that exposure increases liking of musical stimuli.[31–34] Even preferences for basic stimulus properties such as harmonicity have been shown to depend on musical training.[35] This evidence unequivocally demonstrates a key role of learning and culture in preference formation.

Unfortunately, the effect of learning on preference is difficult to control for, or to model. Details on experimental controls are outlined in Section 6.2.9. Modelling is possible due to machine-learning models that approximate musical expectation in humans.[36] However these cannot account for variation amongst individuals, as they all inevitably create one idealized model based on a specific corpus of music.

### Roughness: Rough tones are disliked.

The focal hypothesis of this work has solid foundations in anatomy and physics (Section 3.1), a lot of theory development (Section 4), independent percepts (Section 6.1), the most robust empirical evidence for predictions (Section 6.2), predictions from evolutionary hypotheses (Section 2.2).

Despite the rich history of modelling, a deep investigation has highlighted some fundamental flaws (Section 5.2). There are semantic problems at the heart of the phenomenon due to a potential tautology in the definition of roughness. Modelling approaches have relied on data that is limited, poor-quality, and out-dated, and they often suffer from overfitting. Key assumptions in models have long been acknowledged, yet have not been robustly tested.

### *Harmonicity*: Harmonic tones are preferred.

This hypothesis is perhaps the oldest, although this depends on how you group a set of similar, but potentially independent theories. Like with roughness, predictions from harmonicity models correlate well with preference ratings. Harmonicity is associated with one independent, measurable phenomena – tonal fusion.[37,17] Evolutionary theories have been proposed to explain the origins of harmonicity preferences (Section 2.2).

The biggest issue with harmonicity is with its theoretical foundations and definition. The common definition is that harmonicity is related to the degree of overlap between partials, or how much a sound appears to be 'whole' as opposed to made up of separate, disjoint parts. The conversion from this logical statement to a mathematical definition leaves a lot of freedom of form, so there are many possible models and none are clearly right or wrong *a priori*. For example, one particular model correlates best with preference ratings[38,39,9,24] yet makes two particularly questionable assumptions: It assumes octave equivalence, which may not be universal and innate.[16] It uses one parameter to allow for small deviations from perfect overlap due to uncertainty in pitch processing. The value for this parameter was chosen arbitrarily, and it is assumed to be constant across all frequencies. These models ought to be robustly tested on independent data. Additionally, some have suggested that the success of harmonicity models is an artefact due to collinearity with other variables.[40] The argument is that harmonicity models correlate so well with cultural prevalence that we cannot distinguish whether this success is attributable to one or the other. Similar arguments can be made for roughness models, since they often have overlapping predictions with harmonicity models.[10–12]

*Sharpness*: Tones with energy at high frequencies are disliked.

This hypothesis (or observation) is the least-well developed, yet it contains sufficient supporting evidence to be worthy of consideration.[41,7,20,42–44] It may be that this hypothesis has simply been overlooked because of the historical fixation on harmonicity. Indeed, many studies would miss any effects of sharpness because they typically use stimuli with a small $f_0$ range. One particularly important reason to consider this theory is that its predictions are independent of the other hypotheses. The sharpness hypothesis predicts that tones with high-pitched content will be disliked, while roughness hypothesis predicts the opposite (roughness in general decreases with pitch height) and harmonicity models make no predictions as they only consider frequency ratios between partials. Another benefit of this model is that it is conceptually simple, which leads to well-defined proxies ($f_0$, spectral centroid) and perceptual measures (brightness).

This hypothesis lacks an hypothesis for the ultimate cause of why we would dislike sharp tones. There is also the possibility that preferences for sharpness are a misnomer for preferences for loudness, as perceived loudness can increase with frequency.

| **Theory** | Learning | Roughness | Harmonicity | Sharpness |
|---|---|---|---|---|
| **Foundations** | Statistical learning | Physics, physiology, psychoacoustics | Number theory, psychoacoustics | Psychoacoustics |
| **Modelling** | Information-theoretic predictions trained on a specific corpus of music. | Empirical mapping between f & df and roughness/dissonance for pure tones. Extrapolation to complex tones. | Mathematical models of harmonic overlap, taking into account uncertainty in pitch perception. | Weighted sum of high-frequency content over frequency bands. |
| **Measurables** | Musical training, corpus familiarity | AM rate, AM depth, roughness | Tonal fusion | Pitch height, brightness |

**Table 1**. Summary of theories of proximate causes of auditory preferences.

## 2.2. Ultimate Causes

Preferences may be to some degree innate. This may only be a weak bias towards certain preferences, which can be amplified through cultural evolution.[45] The primary evidence for this bias is that certain musical intervals are more common worldwide than than what would occur purely through chance, although it is possible that the bias is not related to preference.[1,11–14] Similarly, some would point to the similarities in preferences across countries, although this has been demonstrated with a much smaller range of cultures (Section 6.2.9). If there is some innate bias common to all humans, then we can consider three scenarios for how it could arise. Either these preferences are learned from a very young age due to robust, universal environmental stimuli, or else there is something particular about our auditory processing. If the latter is true, then this might have arisen due to evolutionary selection or from other evolutionary processes (neutral evolution, or as a byproduct of selection for other adaptations). Of course, all of these can be simultaneously true to different degrees, but at this early stage of scientific investigation it helps to address the concepts separately. Here I give an overview of these three scenarios, and outline two specific hypotheses, the "Vocal Similarity" hypothesis and the "Roughness Salience" hypothesis.

### Learning during early life

If biases are common to all humans yet not encoded in the process of brain development, then they must have been learned at a young age from universal environmental stimuli. One may enumerate any number of such stimuli, but for an example I will take screaming / crying. Screaming is a high-intensity vocalization that is used by infants as a communicative signal.[46] It follows a form-function relationship since high-intensity increases the range and salience of the signal, thus increasing the likelihood that it will reach the intended audience. Due to the mechanics of our vocal production anatomy, high-intensity vocalizations are rich in non-linear vocal phenomena (e.g. roughness, chaos, etc.).[47] The fact that these vocalizations are used when infants experience negative emotions or pain may lead to negative associations with non-linear vocal phenomena.[48]

### Evolutionary selection of adaptive traits

It is extremely difficult to prove evolutionary hypotheses about selection of complex traits, yet it is still important to consider them. The two hypotheses that have been proposed lead to predictions of similar preferences in other animals, which can be tested. This means they can at least be falsified using methods available to us today, if not directly proven. For the "Roughness Salience" hypothesis, the prediction is that other animals would benefit from perceiving alarm calls in other animals. This would lead to enhanced situational awareness in dangerous environments, which would increase individuals' rates of survival. Thus other animals should exhibit aversion, or at least heightened arousal, towards rough sounds. For the "Vocal Similarity" hypothesis, the prediction is that other animals ought to have evolved specialized auditory processing (and preferences) for vocalizations typical of their species, which would be adaptive through enhancing communication, situational awareness and social interactions.

### Evolutionary byproduct of selection for other traits or drift

Due to the interconnected nature of the brain, it is difficult to infer reasons for historical change, but we can at least try to find whether specific brain circuits and processing capacities lead to biases in perception and preference. For example, the "Vocal Similarity" hypothesis predicts that we have evolved brains that are especially good at perceiving harmonic sounds due to the harmonic nature of human vocalizations, but it does not explicitly say why this would lead to preferences. One such reason could be that we have evolved brains that reward correct predictions (e.g. predictive coding theory). Thus, preferences for certain intervals could be a byproduct of the development of a prediction reward system, built upon the adaptive trait of specialized processing of harmonic sounds.

Another potential region to investigate along these lines is the cross-modal perception of roughness and harmony in relation to tactile and visual aesthetics.[49,50] We name auditory roughness by analogy to tactile perception. Indeed, one can mathematically and perceptually relate amplitude modulation in sound waves to modulation frequencies of surfaces. A sound with no amplitude modulation is not rough, as is a completely flat surface. A small amount of amplitude modulation leads to audible fluctuations, akin to slight curvature in an undulating surface. Increase the frequency of the amplitude modulations, or decrease the periodicity of surface undulations, and you get roughness. Similar cross-modal relations exist between auditory roughness and visual flickering,[51] and between auditory harmony and visual harmony (e.g. symmetry).[52,53] Are these relations purely semantic? Or do they also arise from shared neural circuitry? If roughness is an epiphenomenon of brain development, perhaps neural responses to auditory roughness is general to other domains.

### "Roughness Salience" Hypothesis

In the absence of an established name, we use the term "Roughness Salience" to refer to the hypothesis that rough sounds have become associated with extreme emotional states across the animal kingdom through co-evolution of vocal production and sound processing anatomy. Complementary evidence exists from humans and non-human animals to support the idea that rough vocalizations are regularly associated with extreme emotional states, and are reliably perceived as such. For a detailed summary of this evidence, see [46]. In humans, screams are uniquely rough compared to other vocalizations. Likewise, manmade alarms (e.g. doorbells, buzzers) are rough compared to other manmade sounds.[54] At least for extant humans today, rough sounds appear to be well suited to communication aimed at quickly getting others' attention. This is supported by behavioural studies showing that humans respond faster to rough sounds and are better at localizing/perceiving rough sounds.[55,46] Neurological evidence shows that brains respond qualitatively differently to rough compared to non-rough sounds, activating many regions including the amygdala – a region associated with fear and response to danger.[46,54] In animals, non-linear phenomena are found throughout the animal kingdom and have been linked to alarm/distress calls in mammals. Mammals and non-mammals have been found to respond differently to urgent vs non-urgent calls in animals from other species.[46,56,57]

Since high-intensity vocalizations are useful for increasing range and saliency of communicative signals, and since they reliably lead to non-linear phenomena across animals due to anatomical similarity, it follows that rough sounds are a reliable environmental signal. Thus there may be evolutionary selection of development of specific brain circuits to optimize responses to these signals.

### Vocal Similarity Hypothesis

The vocal similarity hypothesis proposes that we prefer harmonic intervals because they are found in human vocalizations (e.g. speech).[58,59] It cannot be classified according to the above scenarios since it is in principle consistent with all of them. The hypothesis is that human vocalizations are a consistent and important signal that we are exposed to, which we learn to perceive both on evolutionary and developmental timescales. This can potentially lead to a selective pressure on brain evolution to increase our innate ability to process these salient, important sounds. This is insufficient in itself to lead to preferences, but one can integrate this hypothesis with others to supplement this. For example, predictive coding theory (or processing fluency theory) predicts that we gain reward from correct predictions. Hence we may prefer hearing sounds that we have evolved to be proficient at processing. Likewise, one can relate this hypothesis to the "Roughness Salience" hypothesis to link vocalization processing to affect.

## 3. Auditory Roughness

Roughness is hard to precisely define. It is named by analogy to the tactile domain, as how auditory brightness is analogous to the visual domain. Its proximal cause – beats – is better defined. Beats are named in analogy to the rhythmic domain, where beat is a regular pulse. In this case, the beats we are referring to are also a regular pulse, but one described as amplitude modulation (AM), such that one beat is heard for every period of AM, and is described in units of Hz as beat (or AM) rate. At low AM rates, one can clearly perceive the amplitude modulation as a slow wave, which may even be pleasant.[24,60–62] As the rate increases we can no longer hear slow oscillations, and instead hear a fast buzzing sound. This perception of fast amplitude modulation is what we call auditory roughness.[7]

### 3.1. Physical / anatomical basis of beats due to interference

The fundamental basis of the perception of amplitude modulation due to interferences between sine tones lies in a physical / mathematical duality between representations given by the equation

$$\sin(2\pi f_1 t) + \sin(2\pi f_2 t) = 2cos(\pi|f_1 - f_2|t)sin(\pi(f_1 + f_2)t), \qquad (1)$$

where t is time, and $f_1$ and $f_2$ are the frequencies of two tones.[9] The superposition of two sine tones can be described either as a linear superposition of two separate tones (left-hand side of the equation), or as an amplitude modulated tone of frequency $(f_1 + f_2)/2$ with its amplitude modulated at a frequency of $|f_1 – f_2|$ (right-hand side).

Humans hear these sounds as either one representation or the other due to the anatomy of the inner ear.[63] When sound waves reach the basilar membrane, the membrane vibrates in response. The structure of the membrane is tapered, such that different ends of the membrane respond to (resonate at) different frequency ranges. The next step in sound processing involves a discrete number (~3,500) of inner hair cells that are situated along the membrane. Hair cells respond selectively to a range of frequencies, thus acting as bandpass filters. When two frequencies fall within the same filter (i.e. are picked up by the same hair cell) it is difficult to resolve them as separate frequencies. The width of these filters, known as the critical bandwidth, increases with carrier (e.g. mean) frequency, such that it is about 100 Hz at low frequencies, and increases to about 400-500 Hz by 4 kHz. Whether two tones fall within the critical bandwidth is the primary factor that determines whether amplitude modulation will be perceived.

## 4. Roughness Models

There have been about twenty models of roughness (or sensory dissonance) described in the literature.[5,64–82] Despite their numerosity, they can be grouped into two main families: the "idealised tone models" of the sort first proposed by Helmholtz,[5] and the "waveform models" first proposed by Aures.[73] There are a few models that fall outside of this classification that are not considered here;[81,82] for more information see an earlier review.[9] Both families of models share three important steps: transforming frequency into AM, mapping AM to roughness/dissonance, and summation of dissonance over the full AM spectrum. The main difference is the input to the model: idealised tone models use individual frequencies of idealised tones, while waveform models use AM spectra derived from waveforms. Thus idealised tone models are useful for precision testing of interference models and assumptions, since they deal with idealised stimuli that can be carefully controlled, while waveform models are required for processing stimuli with complex waveforms, such as noise or natural vocalisations / sounds. Despite numerous methodological differences between individual models, there is an equivalence between the underlying structures of the families.

### 4.1. Idealised Tone Models

Idealised tone models[5,64–72] originate from *On the sensations of tone*, where Helmholtz noted that interacting tones sounded most dissonant when the AM rate was 33 Hz.[5] Hence, the approach taken by Helmholtz was to "assume a somewhat arbitrary law for the dependence of roughness upon the number of beats" (Chapter X), choosing a function that is zero when there are no beats, reaches a maximum at 33 Hz, and gradually decreases to zero thereafter. Using this function, he calculated dissonance curves for pairs of partials. Helmholtz reflected in Appendix XV that, "Although the theory leaves much to be desired in the matter of exactness, it at least serves to shew [sic] that the theoretical view we have proposed is really capable of explaining such a distribution of dissonances and consonances as actually occurs in nature."

The next century saw three developments that lead us to the present state. First, Plomp & Levelt (and others) measured preference as a function of AM rate and carrier frequency in psychoacoustics experiments, over a range of carrier frequencies.[64] These measurements have been used to fit empirical pairwise dissonance functions to map AM rate to dissonance, replacing the heuristic approach of Helmholtz. Second, it was noted that the AM rate that produces greatest dissonance depends on the carrier frequency, and that beyond this dissonance reduces to a minimum at an AM rate that is lower than the critical bandwidth. Plomp & Levelt noted that the maximum dissonance occurs approximately at AM rates at about a quarter of the CB, and proposed that dissonance is a function of the AM rate relative to the CB. Hutchinson & Knopoff were the first to provide an empirical fit for the critical bandwidth (although they used a different definition of critical bandwidth; see Section 5.2) based on average fundamental frequency of two tones.[66] They used an idealised *pairwise dissonance function* to calculate dissonance for pairs of frequencies. They used a discrete, tabulated function based on Plomp, and algebraic formulae were later developed to replace this.[67] Third, an *overall dissonance function* was developed to combine the relative contributions of different pairs of frequencies in a complex tone. The idealised tone models differ from each other in how they operationalize the pairwise dissonance function (which maps AM rate and frequency to dissonance) and in how they weight the contribution of pairs based on their amplitudes.

### 4.2. Waveform Models

Waveform models[73–80] comprise three main steps: (i) Computation of the amplitude envelope (or some similar construct; for a more comprehensive review see [9]) across frequency bands. (ii) Application of a filter or weights to account for the relative contribution of different AM rates and frequency bands to the sensation of roughness. Some models also add a modulation filterbank stage, where an additional bandpass filter is applied to AM spectra within frequency bands to account for AM masking.[79,83,84] (iii) Summation over dissonance across frequency bands. Step (i) is equivalent to calculating *df* from two frequencies in idealised tone models. Step (ii) plays a similar role to the pairwise dissonance function. Step (iii) is equivalent to the overall dissonance function. Waveform models typically include modeling of the auditory system through a series of filters or transformations in step (i). The added benefit of such complex modelling has not been tested until recently, where it was found that a simple Fourier spectrogram performed equally to (or slightly better than) the more sophisticated models for voice roughness.[80] Waveform models tend to be less transparent about how they choose/optimise the function to map from AM rates to dissonance, in contrast to idealised tone models which all use linear fits to the same sets of data.

## 5. Lessons from Strengths and Weaknesses of Modeling

Models are most reliable when they have concrete foundations and manage to predict a range of phenomena using the fewest parameters. At the most basic level, roughness models are built on a well-tested understanding of the physics of sound and how it is processed by the cochlea. Beyond this, however, cracks appear due to the empirical nature of the mapping between physical sound and psychoacoustic perception. Unfortunately, some models have not been tested on any data,[75,76] or only on the data used to train the model,[79,80] and/or it is unclear exactly what data was used to optimize models.[65,73–75,77,78] Other models have been tested on limited data, using many more parameters than is necessary, likely leading to overfitting. These and other issues have been previously highlighted before,[85] yet here we still are. To help guide future research I will highlight some successes of both types of models and elaborate on some common problems.

## 5.1. Strengths

Collectively, roughness models have been successfully applied to predict subjects' ratings of a wide range of auditory stimuli. Idealised tone models have been successfully fitted to roughness / dissonance ratings of pure-tone stimuli and estimates of the critical band.[64–72] More impressive is how some of these models have been used to predict preferences for dyads and chords composed of complex tones,[9,24,25,27] for the most part without additional fine-tuning (except for the addition of a term for preference of slow beats),[24] including non-trivial predictions of combinations of inharmonic complex tones.[24,25] Waveform models have been used to predict responses to simple stimuli (beating tones, AM tones, AM noise),[73,74,78] engine noise,[77] voice quality,[79,80] and natural sounds.[55]

## 5.2. Weaknesses

### 5.2.1. Tautology and lack of independence in measurements

Roughness is defined as a sensation of fluctuation or buzzing that occurs for AM of roughly 15-300 Hz.[7] Unlike AM alone, which is well-defined both theoretically and perceptually (e.g., we can reliably count beats up to a certain AM rate), roughness is not so strictly defined. How do we know where roughness starts and ends? Helmholtz originally described these tones as "jarring and rough", intertwining the concepts of roughness and dissonance (Chapter 8).[5] Terhardt describes roughness in a similar way, reporting that, "fluctuations are perceived as an unpleasant, disturbing component which usually is called 'roughness'".[86] The first psychological experiments did not ask subjects about roughness, but instead about dissonance.[64,87–89] This would suggest that the discriminating factor between rough and non-rough sounds is dissonance itself, in which case there is an inherent tautology in the definition and measurement of roughness. Eventually researchers pivoted to asking participants about roughness instead of dissonance, to determine maximum roughness,[90] and the boundary at low AM rates between 'fluctuations' and 'roughness'.[91]

We can consider two scenarios. One in which roughness is a percept that can be measured independently of dissonance, and another in which the two phenomena are intrinsically linked in some way. To explain, I present one example as an illustration: the experiment of Terhardt.[91] He presented participants with tones that differ in carrier frequency and AM rate, and asked them whether they heard a "sensation of fluctuation" or a "sensation of roughness". The point where 50% of subjects respond with either answer is the lower limit of the roughness region. What is going through the mind of a subject in this case? Do they have an intuitive response that corresponds to this unique percept called 'roughness'? Or do they determine the boundary based on how dissonant the sound is? Either scenario seems plausible, and thus it is not entirely clear that roughness and dissonance are independent – or how related they are. If roughness is defined as the region where fluctuations start sounding dissonant, then it is tautological to state that roughness is the cause of dissonance.

How do we tell these two scenarios apart? One way to answer this would be to separately measure roughness and dissonance for many individuals. At the level of individuals we can examine to what degree their perception of roughness and dissonance overlap. Comparing individuals, one might expect that roughness is perceived more consistently than dissonance, if auditory roughness is akin to tactile roughness, or loudness – we expect dissonance to vary across individuals since dissonance is at least partly dependent on exposure, and hence should be more variable.

What does it mean if dissonance defines roughness? Then we can conclude that AM is the cause of both sensations. In this case, it is still possible to construct a model of consonance based on AM (or roughness) by doing what idealised tone models do. We measure dissonance for pure tones to get a mapping between AM rates and dissonance, and then use this to construct a model that predicts dissonance for complex tones, either using idealised tone or waveform models. The semantic difference would be that AM rate and depth are factors that determine dissonance, and roughness is simply the name we use to describe that type of dissonance.

#### 5.2.2. Lack of a unified approach

The two types of models are not fundamentally incompatible with each other. Yet no one has tried to integrate the two approaches or understand how and where they differ. Likewise, different models have rarely been tested on stimuli that they were not explicitly designed for. In one example, a waveform model[77] that was designed for predicting the roughness of engine noise was found to perform poorly (Pearsonr's $r = 0.17$) at predicting preference ratings of musical intervals and chords.[9] Altering the model led to better performance ($r = 0.46$), but this altered model was never re-tested on engine noise, so it is not clear if the same model can predict responses to both types of stimuli.

#### 5.2.3. Overfitting

In creating mathematical models, it is a standard heuristic that one should use as few free parameters as necessary. Given enough parameters, it is always possible to fit a model to data. This is captured by a famous, sardonic quote attributed to von Neumann, "With four parameters I can fit an elephant, and with five I can make him wiggle his trunk."[92] How do we know how many free parameters are too many? It depends on how much data, and the complexity of that data. For example, a linear relationship between two variables needs only two parameters. Amongst the idealised tone models, the Hutchinson & Knopoff model stands out as the simplest, having only four free parameters. Two are needed to define a log-linear dependence of the critical band (CB) on carrier frequency, and one is needed to correctly approximate the pairwise dissonance function. Other models are not so concise, for example, Vassilakis' idealised tone model uses eight parameters. In the case of waveform models, things can get very complicated depending on the auditory model, and depending on whether weights of frequency bands are optimized – in some cases, the number of effective free parameters is greater than 10.[77,78] Any model with so many parameters is not likely to generalise outside of its training data, as was seen with Wang's model for engine noise when applied to musical stimuli. The heuristic of simplicity in modelling should not be so easily forgotten or ignored.

#### 5.2.4. Inconsistencies and propagation of errors in the literature

There are some factual errors and inconsistencies in the literature that are quite serious. If I may speculate as to why this occurred, part of the blame is due to inadequate reporting in old papers, such that they are not completely reproducible. Another potential reason is that a lot of the research on roughness was published in German. Whichever the cause, I can point to one inconsistency that has been ignored, and one error that has been propagated through the literature.

Plomp & Levelt calculate dissonance curves, which inspired future idealised tone models. Their calculation involved dividing AM rate by the critical band (CB), but they neglected to state what value they used for the CB. As a reminder, the CB is the point where upon increasing *df*, perception changes from hearing one tone to two. In their original paper, they reported three sets of data in Fig. 9: (i) The CB as estimated by Zwicker from four independent experiments.[93] (ii) Measurements from Cross & Goodwin of the df at which beats can no longer be heard.[94] (iii) Measurements from Mayer of the smallest interval perceived as consonant.[95] In the ensuing works, the most recent data from Zwicker was ignored in favour of the other two sets of measurements, which were used to fit a function for the CB.[66,72] While (ii) and (iii) show similar trends, the (i) is qualitatively and quantitatively different. Given that the CB is a core part of later Helmholtz models, this inconsistency should have been addressed by now, yet it seems to have gone unnoticed (or at least uncorrected). Incidentally, updated estimates of the CB appear to be more in line with the measurements from the earlier two papers.[63]

In the original waveform model paper, Aures produced roughness curves that were calculated from a model.[73] In some studies that followed, authors reported that these calculated roughness curves were actually empirical measurements,[74,77,78] and used them to validate their models (and possibly also to optimize them, but we cannot say due to insufficient detail in methodological reporting).

#### 5.2.5. Quality of empirical evidence underlying models

There is a curious trend of researchers re-using legacy data rather than updating models with more accurate data, including data from over a century ago. The idealised tone models all use the Plomp & Levelt (1965) data for their pairwise dissonance function, which consist of median dissonance ratings from ~10 subjects per datum. The idealised tone models that explicitly model the CB use data from Cross & Goodwin (1 subject) and Mayer (12 subjects) from the 19th century,[94,95] even though Plomp provided updated measurements (20 subjects) in 1968.[90] It stands to reason that we can revise this foundational data in the 21st century, using large sample sizes and modern psychometric methods.

#### 5.2.6. Assumption of Linearity (Additivity)

Models differ in the degree to which the overall dissonance is treated as a linear sum of partial dissonances. Idealised tone models are completely additive. That means, the roughness of two rough sounds heard together is equal to the sum of the roughness of the two sounds heard individually – idealised tone models sum over pairwise dissonances from pairs of partials, weighted by amplitude. For waveform models, roughness is calculated in a similar way to loudness, such that roughness is calculated first within critical bands and then summed across bands. Some models use a cross-correlation term that introduces non-linearity, motivated by work published in a PhD thesis from 1993.[96] More recently, models have also introduced a modulation filterbank stage to account for AM masking,[79,83,84] whereby AM tones can mask other AM tones if they have similar AM rates (see Section 6.1.5).

The assumption of linearity was first noted by authors in early works as speculative, but useful for the purposes of simplification.[64–67,69,70] Some later authors, however, do not comment directly on this assumption.[68,71,72,78,80] One paper concluded that the available evidence suggests that roughness is additive outside of the critical bands (like loudness).[74] This evidence is unfortunately not very convincing, as not only are the conclusions based on few measurements with few subjects,[73,86,97] there are logical inconsistencies in the interpretation of the results. For example, it is reported that, “Terhardt [43] finds that the roughnesses of two different AM tones can be added if they are modulated in phase and if the frequency difference of the carriers exceeds 1 Bark.”[74] However, according to this logic, the results also show that for frequency differences below 1 Bark the roughnesses are superadditive – i.e., the roughness of the combined AM tones is 50% higher than the sum of the roughnesses of the individual AM tones.[86] This seems unlikely, so it is unwise to take these results as strong support for or against the assumption.

Indirect evidence from consonance modelling of dyads and chords leads to divergent conclusions. In models where the number of tones is included as a predictor variable – if the number of tones affects consonance, then consonance is not additive – one study found an effect,[9] while another found no effect.[27] Another study found that consonance of chords could be predicted by the consonance of dyads, which implies additivity.[98]

# 6. Experiments

## 6.1. Humans: AM Perception

### 6.1.1. AM rate

Studies on AM rate include those that asked participants to explicitly count beats, and more often those that measured just-noticeable-differences (JNDs) that describe our ability to discriminate between rates of different magnitudes. Low AM rates can be reliably counted,[99] even when presented dichotically.[100] Humans can discriminate between rates at a resolution as low as ~0.5 Hz, although JNDs increase with AM rate, up to 25 Hz for AM rates of 320 Hz.[101,102] Performance is similar for sinusoidal and pseudorandom modulation envelopes.[103] One paper studied the effect of carrier frequency on AM rate discrimination, but it is unclear if there is an effect given the measurement variance and small sample size.[101] The implications of this for roughness perception is that at high AM rates, perceived roughness should vary with AM rate slower than at low AM rates.

### 6.1.2. AM depth

What is the absolute threshold for modulation detection in terms of modulation depth? And how does it depend on carrier frequency and AM rate? It can be difficult to compare studies, since the threshold depends on sound pressure level (SPL) also, and this is not consistent in experiments. Also, some studies report discriminability rather than absolute thresholds, so they cannot be directly compared. However we can summarize the results as follows: Within the critical band the threshold appears to be independent of AM rate up to 100 Hz – somewhere in the region of 2-5% modulation depth.[104–108] After 100 Hz, the threshold starts to increase at a rate of about 8 dB per octave.[107] Note that studies tend to show a non-monotonic dependence on AM rate – performance is flat, then drops, and then increases. This last increase can be attributed to discrimination on the basis of frequency difference (sideband) detection as the AM rate crosses the critical band. A separate review reported that discrimination is peaked at 4 Hz,[109] however this is based on data from 4 subjects in 1952 reported without errorbars.[104]

How well can people discriminate between different levels of modulation depth? JNDs are strongly dependent on the modulation depth, with smaller differences being perceivable for low depths.[110–112] E.g., for depths of 5% differences of 5% are detectable, while for depths of 50% differences of 25% are detectable. The implications of this for roughness perception is that at low AM depths, perceived roughness should vary with AM depth slower than at low AM depths.

### 6.1.3. Phase effects and envelope asymmetry

An investigation by Pressnitzer and McAdams (1999) used an innovative methodology and produced novel findings that have, in my opinion, not been given sufficient attention.[113] They varied the relative phase of sine tones so that when the phases are equal or when they differ by π they act as normal AM tones, but in between this range the AM depth is attenuated. Thus, this is a method of systematically varying the AM depth while keeping the frequency and AM spectra constant. They showed significant effects of the relative phase of interacting tones, and surprisingly that there is an effect of temporal envelope asymmetry on roughness perception. Phase effects thus ought to be taken into account in roughness modelling, and more investigations are needed to understand the effect of envelope shape in addition to rate and depth.

### 6.1.4. AM quality: fluctuation and roughness

Perception of AM is described as having two distinct qualities. At low AM rates, a sensation of fluctuation, and at higher AM rates a sensation of roughness. There are two basic questions that arise. What is the boundary between the sensation of fluctuation and the sensation of roughness? Then, considering that fluctuation and roughness are not simply perceptual categories, but continua with magnitudes, what is the meaning of the magnitude of fluctuation and roughness? I will keep this section brief, since a lot is covered in other sections due to semantic overlap between roughness and consonance. See Sections 5.2.1 and 6.2.1 for more on this semantic overlap. Section 6.2 covers the broader topic of consonance measurements, including some studies where roughness and consonance are not distinguished and some studies that report measuring roughness. To summarize briefly, roughness and consonance have been at times defined tautologically, so it is not quite clear what exactly the percept of roughness is. One scenario is that the boundary between roughness is the point where AM becomes unpleasant, and the magnitude of roughness is the degree of unpleasantness.

Fluctuation strength has been studied much less than roughness, but it has been measured[7,114–116] and modelled[7,117,118]. In contrast to roughness, fluctuation strength was found to not depend on carrier frequency in the range of 125 Hz to 8 kHz. Fluctuation strength was found to be maximal at 4 Hz, and to have minima at 0.25 and 32 Hz, for AM sine tones, frequency-modulated (FM) sine tones and AM noise. A monotonic relation was found between fluctuation strength and modulation depth (and SPL). Since roughness magnitude depends a lot on carrier frequency as well as AM rate, there are some curious observations when comparing fluctuation strength and roughness. At low carrier frequencies, the point of maximum fluctuation overlaps with maximum roughness – e.g., one study found maximal roughness at 5 Hz AM rate for a 125 Hz carrier.[64] As you increase the carrier frequency, the AM rate that produces maximal roughness increases, as does the point at which sounds start to sound rough. At sufficiently high carrier frequencies (e.g., 10 kHz) the idealized tone models fitted to Plomp data eventually predict that there is a region at about 30 Hz where sounds produce negligible roughness and fluctuation, which seems odd – what is the sensation of a 10 kHz sine tone with AM rate of 30 Hz if it is not rough nor fluctuating? What strategy do people use to decide on the evaluation of fluctuation strength? Ultimately, there have been very few published measurements of fluctuation strength of AM, all by the same research group, and the phenomenon is not precisely defined. Thus there is more work to be done to establish what is meant by fluctuation strength, and both the nature and location of the boundary between fluctuation and roughness. This is supported by non-peer-reviewed results from a separate group that found that subjects' estimation of fluctuation strength for FM tones did not conform to previous results.[119] Instead, they found an extreme lack of consistency between participants' responses. It is worth considering the possibility that fluctuation is simply a loose description, and that AM rate does not lead to a reliable difference in perceived fluctuation strength that is shared across humans, and that there is no sharp boundary between fluctuation and roughness.

### 6.1.5. AM masking – the modulation filterbank concept

Frequency masking occurs when multiple frequencies fall within the same critical band. This not only leads to the sensation of AM (Section 3.1), but also affects loudness perception. If you have two sine tones with the same amplitude, the perceived loudness is greater if they fall in separate bands. A similar phenomenon has been observed for AM perception. Frequency masking is understood to originate from the structure of the cochlea, whereas AM masking is likely due to downstream neural circuits. However, the details have not been fully resolved and the main evidence in humans comes from psychoacoustic studies, which supports the concept of an effective modulation filterbank. That said, the evidence is quite consistent, although the width of the filter is not precisely known. Studies have used multiple experimental methods on both noise and sine tones, showing clear evidence for AM masking.[83,84,120–123]

## 6.2 Humans: Consonance / Preference

### 6.2.1. What is measured?

In Section 5.1 I pointed out the difficulty of defining roughness, as it is a subjective percept, and is often defined in reference to consonance. We run into similar issues – subjectivity, circularity – with the definition of consonance. One common, simple definition of consonance is pleasantness.[124] Other definitions include not only this concept of pleasantness, but also the hypothesized cause. For example, the Grove dictionary describes dissonance as, "a discordant sounding together of two or more notes perceived as having 'roughness' or 'tonal tension.'",[125] invoking theories of roughness and musical expectation. A more instructive place to look for the definition of consonance is in the studies where it is measured, since this is the source of our empirical understanding of consonance. Studies have described consonance to subjects using a plethora of terms including: consonance, musical fit, smoothness, pleasantness, beauty, attractiveness, harmony, clarity/turbidity, valence. Is there one correct description? Do all descriptions point to the same underlying phenomenon? Are there multiple relevant dimensions that are reached by using questions with different terms? Do people use the same mapping between words and underlying perceptual dimensions?

Let's start by thoughtfully considering the basic practice in psychology of asking people questions to infer something about their mental representations. If you were to ask someone to arrange a set of musical intervals in terms of how "round" they sound,[126] what would they do? If they can reliably order the items across multiple tests, then we can be sure that they have at least found a strategy of mapping the term "round" to a reliable perceptual continuum. Is this percept unique to the word "round" or is it something more general? Given a set of musical intervals, how many reliable mental representations do humans have? There are objective quantities like pitch height, interval size, and AM rate/depth. Then there are quantities like loudness that have a clear physical origin – sound-pressure level – which is non-linearly transformed to a subjective quantity. And then there are qualities like pleasantness or tension that are purely subjective – we may hypothesize about objective origins, but the definition of pleasantness itself is entirely centered on the subject and their pleasure. We can in principle ask subjects about any number of quantities/qualities, but we must be careful about interpreting such results.

Although there are certainly many consistencies across studies – consonance ratings are often correlated even when probed by different wordings – I first highlight some discrepancies within the literature where studies have tried to probe more than one perceptual dimension. In two parallel highly-powered studies (410 and 281 subjects), Lahdelma & Eerola asked subjects to rate chords along 5 and 9 dimensions, of which "preference", "valence", "energy" and "tension" were included in both studies.[127,128] Some of the ratings led to reliable correlations across the two studies: tension and energy ($r = 0.78$ and $r = 0.5$), valence and preference ($r = 0.63$ and $r = 0.49$). Some did not: valence and energy ($r = -0.27$ and $r = 0.48$), valence and tension ($r = -0.78$ and $r = 0.08$). In another study subjects were asked about consonance, pleasantness, stability and relaxation.[129] Given that consonance is often defined as pleasantness, it is surprising that they found a low correlation ($r = 0.32$) between these two dimensions. Thus it is not clear how reliably these terms are mapped to perceptual continua.

To understand how many perceptual dimensions underlie subject responses, one can use dimensionality reduction techniques like principle component analysis (PCA). Topken et al. measured 18 dimensions, and found that 4 PCA components were sufficient to explain 50% of the variance in responses.[62] Unfortunately it is difficult to interpret these dimensions, since the authors only reported half of the weightings of the components. At the very least this suggests that the number of perceptual continua is considerably lower than 18. An early study looked at 10 dimensions, including words like "wide" and "round".[126] I used PCA to analyse these results, finding that 77% of the variance can be explained by a single dimension. Similarly, Lahdelma & Eerola found very high correlations ($r > 0.9$ for musicians, $r > 0.66$ for non-musicians) between all 7 dimensions studied, suggesting that participants related each term to a single underlying perceptual continuum.[130] Two other studies looked at ratings of consonance with, respectively, valence and roughness.[20,131] I calculated the correlation between mean ratings per chord using publicly available data (data from Fig. 1 in [20], and an OSF repository[131]), finding strong correlations between consonance and valence ($r = 0.87$), and roughness ($r = 0.95$).

Sometimes related concepts are used that are not always clearly linked to consonance by definition. For example, several studies have asked participants to rate how 'finished' a note sounds, which is related to the music-theoretic concept of tonal stability.[19,132] Although the authors did not link this percept to consonance, other studies have shown correlations between stability and consonance ratings. Likewise, emotion is often studied as a separate entity from consonance, but valence and consonance are often found to be correlated.[133] A somewhat tangential set of studies looked at the effect of musical intervals on responses in affective priming experiments – the idea is to present words alongside intervals, and if the valence is matched then this leads to faster reaction times in evaluating the valency of the words.[134–136] These offer some evidence to consider, but it is perhaps less direct than measuring consonance ratings.

Overall, some papers point to the idea that there is only one dimension that is reliably invoked to answer questions about semantically distinct categories. Some papers suggest that there are multiple dimensions. Correlations between ratings have been found to be consistent within studies, but have opposite or no correlations when comparing across studies. It is clear that there is still work to do at the fundamental level, to really understand what we are measuring. Future work in this area could include interview-based approaches as has been done for timbre,[137] or massive sampling and dimensional reduction as has been done for emotional responses to various stimuli.[138] One particularly useful methodological control that has been employed in cross-cultural research is the use of other stimuli that are thought to be universally recognized – happy vs angry faces, laughs vs gasps.[15,139] These are used to show that the words used are reliably interpreted, and perhaps they should be used in all studies, not just cross-cultural ones.

### 6.2.2. How is it measured?

There are three types of methodology used to assess preferences for / consonance of tones: ratings, comparisons, and manipulation of stimuli by subjects. The most common approach is to ask people to rate sounds. This is usually on an ordinal scale with about 4 to 11 points, although some have also used sliders to get finer-grained ratings, and some have used binary ratings. The most common alternative is to ask subjects to compare pairs of tones. Typically this is a binary comparison, but some have also used an ordinal difference rating. Finally, some of the oldest studies used continuous stimulus exploration by allowing participants to adjust a knob on a machine to optimize perceived consonances. More recently an innovative online method called Gibbs sampling with people (GSP) has been used to explore dense stimulus space. Like the method of adjustment, GSP allows participants to explore a part of the stimulus space using a slider to optimize "pleasantness".[28] More importantly, this method samples the most relevant parts of stimulus space by chaining participant responses so that one person's optimal point will be used as a starting point for the next participant. Given the ethical and financial costs of psychological experiments, it would make a lot of sense to compare these methods. Future work should identify the most efficient and reliable way of measuring preferences to minimize cost and measurement variance.

### 6.2.3. Stimuli: small intervals, pure tones

It is instructive to first look at evidence for preferences that are most closely related to the roughness hypothesis compared to other hypotheses – dissonance elicited by pure tones. I need to restate, however, that this is perhaps not a prediction of the roughness hypothesis, but rather the basis of it (see Section 5.2.1). Nevertheless, none of the other theories can account for this behaviour: aversion to pure tone dyads with frequency differences in the range 15 - 100 Hz, and the scaling of this range on carrier frequency. For more detail see Section 2.1. Here is a summary of the important points.

Sharpness is based on absolute, not relative pitch, which conclusively discounts it. Harmonicity models differ in their predictions, but none reproduce the typical dissonance curve for pure tones, or the scaling with carrier frequency. As for cultural learning, one could argue that the lack of minor second harmonies in music is why they are disliked, but this doesn't explain dissonance over the full range of AM rates (i.e., intervals that are not found in music), nor does it explain the scaling. The neurodynamic hypothesis appears to be a promising candidate for replicating these results, but it needs more development to assess how well it predicts dissonances over the full range of AM rates or scaling. For example, a model showed that semitones do not lead to stable phase-locking, but only for a few small-integer-ratio intervals.[140] A more general model showed that in principle one can calculate phase-locking stability for irrational numbers, but has not been tested on preference data.[8]

Studies with pure tone stimuli and small intervals are thus the most informative about roughness. Humans consistently dislike small intervals or intervals with AM rates in the range 15-100 Hz. This has been found in almost most studies, including in populations that have little exposure with Western music. See Section 6.2.9 for the few exceptions discussed in detail.

### 6.2.4. Stimuli: dyads, chords and sequences

It becomes increasingly difficult to determine how roughness relates to preferences as stimuli become more complex due to the number of notes in a dyad/chord, and the number of dyads/chords in a sequence. There are several problems with complex stimuli. For many stimuli – e.g., small-integer-ratio dyads of complex tones – the predictions of roughness, harmonicity, cultural learning and neurodynamics all overlap considerably. For chords with many notes we run into unresolved issues with the assumption of additivity (see Section 5.2.6). None of the hypotheses make specific predictions about sequences of chords, as opposed to single chords.

Studies using dyads are most useful when they use intervals that lead to differential predictions by different hypotheses. One way is to use intervals that are not common in global music – i.e., 12-tone equal temperament (12TET). Studies using so-called "dense ratings", which sample uniformly over interval space are most informative,[24–26] followed by studies that use the Bohlen-Pierce scale which only partially overlaps with 12TET intervals,[21,20,22,23] and other studies that investigate non-12TET intervals.[61,62,64,89,141–143]

### 6.2.5. Stimuli: timbre

Timbre is especially interesting, since the predictions of roughness models can depend a lot on timbre. However this complicates evaluation of results across studies, since they mostly use stimuli with different timbres. There are 4 typical timbres that are used, which we can classify as (i) pure tones, (ii) idealised tones, (iii) synthetic instruments, and (iv) real instruments / voices; these can all potentially be harmonic or inharmonic, although the majority of the time harmonic tones are used. I note in brackets the number of studies in which the timbres have been used. Some studies used multiple timbres.

(i) Pure tones are of course a limiting case where sounds consist of a single frequency (8 studies), and are most common in early studies. (ii) Idealised tones are created from specifying the partials that make up a complex tone, and their relative amplitude (and potentially phase), and adding them together (13 studies). The most common way of creating these is to specify the number of partials in a harmonic series, and to apply harmonic roll-off to steadily decrease the amplitudes of partials by some decibels (10 studies). Other approaches include using only even or odd harmonics, or to only include some arbitrary set of harmonics (3 studies). Inharmonic tones are used less often (4 studies). Inharmonic tones have been generated systematically by randomly jittering the frequencies of partials, and by compressing or stretching – adding or subtracting a constant factor from the frequencies of partials. In principle, any set of partials can be combined to create an inharmonic tone. One such example is the creation of an idealised tone based on measurements of the inharmonic spectra of Indonesian instruments. (iii) Instrumental timbres are the most common in recent studies, and are typically made with some sort of digital audio workstation (17 studies). While piano is the most common choice of instrument, others include strings, voice, organ, clarinet, saxophone, sitar. (iv) Studies using real instruments or voices are the fewest (5 studies). Recordings of single notes or chords can be further manipulated by the use of pitch shifting or spectral manipulations. Out of possible timbres, the best choice for disambiguating theories are inharmonic timbres that can lead to very different predictions. Next best is to use harmonic timbres with carefully selected partials and partial amplitudes. Varying timbre by instrument is not particularly informative since they do not lead to large differences in predictions – except in cases where the instrument is partly or wholly inharmonic.

Timbral manipulations lead to predictions of atypical consonance profiles. Many of these predictions have been found to be correct. For example, a study used real sounds from a carillon that has a timbre with both harmonic and inharmonic partials.[25] The authors extracted an idealized timbre, which was used to make specific predictions from a roughness model. This model successfully predicted a switch in preferences from a major third to a minor third when a carillon timbre was used instead of a purely harmonic timbre. Another recent study also found that roughness models could predict some of the non-intuitive changes to preferences that resulted from multiple types of timbral manipulations.[144] Although these studies managed to disambiguate psychoacoustic from cultural effects, a harmonicity model was also found to successfully predict many novel findings, which means we lack the clarity needed for causal inference. Ultimately, due to the impact of timbre on preferences and the broad range of timbre used in studies, differences in effects across studies can be partly due to choice of timbre. For comparative testing of preferences in different populations, the use of standardized timbres is advised.

### 6.2.6. Stimuli: diotic vs dichotic

Stimuli can be presented the same to each ear (diotic) or differently across ears (dichotic). Since amplitude modulation is caused by physical interference of waves, presenting two pure tones to different ears should not lead to any beating or roughness. Surprisingly this is not what happens, as we still hear beats but at a much attenuated level. This phenomenon is called binaural beats. Beats arise in this case due to interactions at a neural level. The possibility of interaural interference can be ruled out based on measurements of interaural attenuation in earphones,[145] and from cochlear measurements in chinchilla.[146]

How does perception of binaural beats compare to monaural beats? Some old studies found that binaural beat detection was in general reserved to carrier tones below 1 kHz, and for modulation rates lower than 100 Hz.[147–149] Surprisingly, these findings have not been extended or replicated much,[150] so more work is needed to definitively establish the limits of perception for binaural beats. On the strength of perception, one found that binaural beat strength was matched by monaural beating tones at a small modulation depth, close to the absolute threshold for monaural beats.[100] Likewise, an EEG study managed to observe signatures of beats in dichotic tones with carrier frequency of 400 Hz and AM rate of 40 Hz. This signature disappeared for high carrier frequency (3,200 Hz) and for low volume (30 dB), while the signature persisted in these stimuli when presented monaurally.[151] Thus, binaural beats are perceived much weaker than monaural beats, across a much more limited range of stimulus space.

Given this evidence, the use of dichotic presentation is an excellent way to control for roughness when evaluating consonance. It is then surprising that this approach has to date only been used in five studies on consonance.[152,35,153,15,154,155,23] Five of these involved comparison of responses to diotically and dichotically presented stimuli, and in each case diotic presentation led to stronger responses, indicating a role of roughness.[35,153,15,155,23] Conversely, dichotically-presented stimuli also lead to clear preferences, which is one of the primary pieces of evidence that shows that roughness is not the only factor underlying consonance.[152]

### 6.2.7. Stimuli: natural and manmade sounds

Complex natural and manmade stimuli are important to study as ecologically relevant signals that are complementary to musical stimuli. They are generally much more complex and diverse than synthetic tones and musical stimuli, which makes them ill-suited for investigating the fundamentals of roughness perception and preference. Yet, it is important to note the varied cases where roughness (or pleasantness) has been estimated. These include vehicle noise,[77,156,157] human voice (speech, song and non-verbal vocalizations),[158,79,159,160] and a variety of natural and manmade sounds.[161,42] Some have also proposed to evaluate medical alarms in terms of roughness.[162] Developing a precise theory of roughness is important for understanding our perception of these sounds.

### 6.2.8. Individual differences

The typical approach when studying consonance is to report summary statistics across participants for each test item. Very few studies have investigated individual differences. Despite the lack of studies, the four that exist all show considerable differences between individuals. Furthermore these studies are not overlapping in terms of stimuli, so it is difficult to tell how much agreement there is between them. Here I review some of the key features of each study.

(i) The oldest study on individual differences studied the pleasantness of dyads and chords, and used methods to estimate individual differences in roughness and harmonicity preference.[35] They found that differences in preferences for harmonicity, but not roughness, correlated with both musicianship and preferences for dyads and chords. Preferences for roughness were determined by the difference in individuals' preferences for diotic vs dichotic intervals (approximately semitones). The results of individual differences in preferences for roughness were not explicitly shown, so limited inferences can be drawn.

(ii) Another study looked at preferences for dyads and chords in 12TET.[27] They used a method to separate measurement variance from individual variance, and found that semitones exhibited the highest level of differences in preferences.

(iii) The third study is the most interesting, as responses to non-musical AM tones were recorded over a range of 10 - 250 Hz, and the results showed two separate populations of people who responded differently.[163] One population responded in the manner found in previous studies, where aversion peaked at

~40 Hz. A separate population did not experience this peak, and instead aversion continued to increase with AM rate across the entire range.

(iv) Finally, one study measured roughness ratings across a large range of carrier frequencies and AM rates for 7 subjects.[142] Although this study mainly analysed group averages, they also plotted individual results for the beat rate at which roughness was maximal, as a function of carrier frequency. This data shows wide disparities between individuals. Of particular note are the differences for 63 Hz carrier frequency, where most reported maximal roughness as AM rates of about 20 Hz, while for others this was 60-80 Hz.

This small, but interesting set of evidence leaves a few questions. Study (i) leaves a lot unknown, because few stimuli were used for roughness preferences, and raw results of individual differences were not reported in great detail. Study (ii) suggests that, despite semitones being the standard interval that is most reliably disliked across cultures, it is the interval with the most diverse responses within a culture. Study (iii) suggests that the differences in preferences for AM stimuli are not only large, but that they are potentially bimodally distributed within a culture. This is consistent with evidence from study (iv), although this study had few participants and did not report all of the raw data. One caveat about study (iii) is that perhaps the qualitatively different responses arise from different strategies for responding / different ways of interpreting how to respond to questions, instead of different psychoacoustic responses. At the very least, these studies suggest that there is much more to be learned from studying individual differences within cultures, and that we have been blinded by the common practice of mainly studying average responses of groups. How can semitones be both the most reliably disliked across cultures, yet the least reliably disliked within cultures?

#### 6.2.9. Within and across cultures

If two populations are exposed to different musical traditions and if preferences are affected by cultural exposure, then the populations should exhibit different preferences. There are three issues with this approach in practice. (i) What needs to be different across musical traditions in order to alter preferences for musical intervals? Plausible candidates are the tuning systems, and the tonal hierarchies used. Most societies that have been studied use similar tuning systems – differences between equal temperament and similar tuning systems (e.g. Pythagorean, just intonation) are so small as to be for the most part negligible.[13] The most prevalent scales and their tonal hierarchies do differ, although they often retain salient features such as octave equivalence and a predominance of fifths.[164] Furthermore, there is evidence of increasing homogenization of tuning systems, as the world plausibly moves towards 12TET as a standard tuning. (ii) Globalization of music streaming has blurred the musical boundaries between countries. Especially in cases where subjects are drawn from the student population, we may expect musical tastes to overlap across countries. (iii) Musical tastes vary substantially within populations. The tonal preferences for someone who listens to pop music may differ significantly from others who listen primarily to jazz (which uses rich tonal palettes) or metal (which often uses heavy distortion and harmonies that are typically considered dissonant). Thus, cross-cultural studies are potentially confounded by the use of similar tuning systems, overlap between musical tastes, and heterogeneity within cultures.

The most consistent finding across cultures is that humans consistently dislike small intervals or intervals with AM rates in the range 15-100 Hz. This has been found in the vast majority of studies,[165,126,64,166,89,167,61,35,143,168–170,20,154,22,171,24,25,27,26] including in populations that have little exposure with Western music. There are two exceptions for studies which asked about pleasantness or preference: A study in Nepal had mixed results, but it used small sample sizes (6 & 10).[172] A study in Greece found no significant difference between semitones and other intervals – however the fact that they didn't find *any* differences between 15 intervals may indicate that their unconventional methodological choice ('yes' / 'no' preference ratings) is unsuited for estimating consonance ratings.[44] There are three exceptions for studies that asked about the "need for resolution", "tension" or "finishedness". In one study in India, semitones and whole tones were not rated as especially "restful".[132] In a recent replicate, study semitones and whole tones were found to be tense by musicians, but not non-musicians.[173] Another study in Papua New Guinea found that the group with least exposure to musical harmony reported no difference in "finishedness" for semitones compared to other dyads.[19] It is not clear how much or how reliably the concept of "tension" or "finishedness" matches up with pleasantness, so these exceptions do not directly contradict other findings of aversion to semitones.

Preferences for other intervals and chords are overall much less consistent than for semitones. Some studies find high correlations in mean pleasantness of intervals and chords across countries: USA vs Japan ($r >= 0.88$);[166] one study using subjects from 10 countries found a mean correlation of r ~ 0.7.[26] But in general a mixed picture emerges from cross-cultural comparisons: Pakistan vs UK;[133,139] Canada vs India;[132] UK vs India;[173] Italy vs Nepal;[172] USA vs Bolivia;[15,174] Germany vs Cameroon;[175] Australia vs Papua New Guinea.[18,19] To give one specific example, the study with subjects from 10 countries, when broken down into pairwise comparisons by country, reveals significant differences that are obscured by the mean correlation: responses from Mexico are uncorrelated with Indian responses ($r = -0.03$) and only weakly correlated with Nigeria ($r = 0.24$). Mexican participants rated tritones as more pleasant than fifths, in contrast to what one would expect based on Western musical theory. Among the countries with high correlations there are Germany and USA ($r = 0.93$) which have a shared musical history and tonality, but also Poland and Turkey ($r = 0.88$) which do not. It remains unclear how much one can interpret these results, given that we don't know to what degree sampling participants from different countries really controls for cultural influences on consonance.

What about within-culture consistency? Many repeated measurements of the standard dyads in the chromatic scale lead to similar ratings.[166,170,27] However, Lee et al. (2025) studied intervals uniformly distributed rather than at chromatic scale points, and found low split-half correlations within countries, despite having ~300-500 participants per country.[26] In a rather spectacular set of results, one study in Germany found the opposite trend that is typically reported – out of a set of 15 dyads ranging from 4.3 to 5.5 semitones, the fourth (interval with a ratio of 4/3) was found to be the least pleasant (as were other small-integer ratio intervals).[62] Studies consistently find strong effects of musical training, establishing one clear cause of within-population variation.[35,15,131,130,173] It also seems likely that there may also be an effect of which musical style one plays, but this has not been systematically explored. The fact that there is often significant diversity in participants' responses within-societies indicates that random (or structured) within-population variation can be a driver for differences across populations. Likewise, the fact that musical training can lead to convergence in consonance ratings, along with the prevalence of musical training in globalized classical and popular music, suggests that this could be a driver for similar ratings across populations.

## 6.3 Animals

Since roughness is an intangible percept that arises in our minds, we cannot directly compare the sensation that humans and animals feel. However, given that roughness is effectively defined as the range of AM rates that are disliked, we can investigate whether there is an equivalent range in animals. We first discuss the evolutionary and developmental hypotheses (Section 3.2) that could lead to roughness aversion, to figure out what ought to be the most promising future steps to take. Then we highlight the sparse examples of experimental evidence for AM preferences in animals.

### 6.3.1. Which animals should experience roughness aversion?

If roughness aversion is an epiphenomenon of the development of certain auditory perception capacities, we should expect to find roughness aversion in animals with similar capacities. Which capacities? Pitch and AM perception are obligatory. But due to the convoluted nature of biological development, we do not know which other specific capacities could lead to AM aversion as a byproduct. In lieu of knowing these capacities, a prudent first step is to study animals whose hearing capacities overlap with humans' capacities in general – e.g., birds. If roughness aversion is an evolved response, developed through exposure to high-intensity alarm/distress calls in the environment, then we should expect to find roughness aversion in animals with the same environments, and the same evolutionary pressure to respond to these environments. Unfortunately the two key components here are not very well defined, and open to a colorful canvas of speculative proposals. The bottom line is that given the hypothesis, we could expect to find roughness aversion in many other animals. Out of the several unknowns – animal AM aversion, capacities, environment, and evolutionary pressure – in this framing, the ones that are easier to unpick are the preferences of animals themselves. It is difficult to even pin down the concepts of environment and evolutionary pressure, let alone define and control for / measure them. As we shall discuss later, it is relatively straightforward to determine preferences for animals using standard experimental paradigms.

### 6.3.2. What AM rates should be tested?

The first prerequisite for inferring AM as a cause of preferences is to determine that animals perceive AM. From a physiology standpoint, animal hearing is somewhat similar to humans. AM will only be perceived when tones interact within frequency bands, defined by the critical band (CB). Beyond this, perception depends on neural tuning to AM.[176] Critical bands have been determined experimentally for a large range of animals – e.g., birds,[177–179] rodents,[180,181] cats(Pickles),[182] primates,[183,184] etc. Neural tuning to AM rates has been also determined for a smaller range of animals.[185–189] For animals where we have this information, we can limit testing of AM rates to the ranges where AM is perceived. To narrow down the range further we can make two types of assumptions: We can reason by analogy to humans that AM aversion will occur at some fraction of the CB – i.e., that aversion will be most severe when AM rates are about 20-40% of the CB. Another assumption is that animals should experience aversion to the same AM rates as in humans – i.e., 15 - 100 Hz. The latter assumption fits logically with the evolutionary hypothesis. One can deduce this since the proportion of environmental signals associated with distress is independent of the physiology of different animals, if we assume that environmental signals are pooled across many animals. Following the same logic, the former assumption is more consistent with the developmental hypothesis, since the AM rates that elicit maximum aversion scales with the CB (i.e., it is not constant). Thus the conservative choice appears most prudent – study all AM rates that are perceived by an animal.

### 6.3.2. Experimental evidence for AM aversion/preference in animals

There are many studies on discrimination and perception of AM or musical stimuli in animals. These are important for establishing the possibility of AM aversion, but they do not offer any direct evidence of AM aversion. The fact that a rat can discriminate between 'consonant' vs 'dissonant' intervals[190] does not explain about what information (e.g. 'consonance', AM strength, or some other salient feature) was used in discrimination, and this tells us nothing about their preferences. For more information on this topic, see [1,190]. There are only eight studies on preference[191–199] They show mixed support for consonance/dissonance preferences, and the evidence is not particularly strong – sample sizes are small (1, 5, 6, 12, 16, 39, 40, 81), and the studies have never been replicated. For the purpose of testing AM aversion, the ideal stimuli would be AM tones of different rates. Unfortunately most of the stimuli chosen are not ideal for testing AM aversion. The stimuli range from having no AM tones[196] complicated stimuli instead of simple, including melodic sequences[193,194,197] and chords (budgies). Those studies that looked at simple musical dyads only used 1 or 2 examples of dissonant and consonant dyads[191,192,195,197]. One study on mice merits a detailed discussion, as it stands out for the breadth of AM stimuli tested, and the strength and robustness of its surprising results.[198]

Postal et al. studied the behaviour of 39 mice housed in an enclosure with a separate chamber with water. When a mouse entered the chamber, they were either presented with a stimulus or silence. In one protocol only one type of stimulus was presented at a time, so animals did not have a choice of stimuli. Four separate types of measurements were made. These measurements show similar results, so I will only mention one – how long the mouse stayed in the chamber. In this protocol, the four stimuli were complex dyads: octave, fifth, major seventh and tritone. These stimuli all led to mice staying in the chamber for considerably less time than silence, and there were no differences between the stimuli. The next set of stimuli used were AM tones or noise with rates of 2, 5, 15, 25, 35, 70, and 140 Hz, including silence and tones/noise without AM as controls. Once again, the mice preferred silence to sound. However, comparing across AM rates, the mice showed the exact opposite of what is observed in humans – they preferred rates of 15 - 70 Hz over faster or slower AM rates, or no AM. These results were further corroborated by a second experimental protocol where mice were offered two choices.

We must be careful about jumping to conclusions on the basis of one strong study. Nevertheless, what are the potential implications of this? First it would seem to contradict the idea that human AM aversion is due to environmental sounds, since we observe opposite behaviour in mice. To save the evolutionary hypothesis, one would have to develop the ideas about environment and selection pressure to the point that it can explain the differential behaviour across species that plausibly share the same environment. Second, what about the developmental hypothesis? In humans we observe AM aversion at around 25% of the CB. Postal et al. used only a single fundamental frequency of 4 kHz, so we know nothing about how AM aversion scales with carrier frequency in mice. We do know, however, that the mouse CB is considerably higher than humans – for a carrier frequency of 4 kHz the CB is about 3 kHz (compared to ~500 Hz in Humans)[180] Thus, when viewing the AM rates in terms of fractions of the CB, they are less than 5% – for mice, these are slow AM rates. This leads to a curious, yet speculative, similarity in humans – there is evidence that humans actually prefer slow AM rates to no AM.[24] It remains to be seen how mice will respond to AM rates that span the whole range of what they can perceive.

In summary, there is very little data, most of it is unreliable, and even the best study has its caveats. Testing between evolutionary and developmental hypotheses is in its infancy. We strongly encourage replications of Postal et al., albeit with additional care taken to account for the specific auditory capabilities of the animal to be studied.